# TAnet: A New Temporal Attention Network for EEG-based Auditory Spatial Attention Decoding with a Short Decision Window

Yuting Ding
Department of Electronic and Electrical Engineering
Southern University of Science and Technology, Shenzhen, China
12332166@mail.sustech.edu.cn

Fei Chen
Department of Electronic and Electrical Engineering
Southern University of Science and Technology, Shenzhen, China
fchen@sustech.edu.cn

*Abstract*—Auditory spatial attention detection (ASAD) is used to determine the direction of a listener's attention to a speaker by analyzing her/his electroencephalographic (EEG) signals. This study aimed to further improve the performance of ASAD with a short decision window (i.e., <1 s) rather than with long decision windows ranging from 1 to 5 seconds in previous studies. An end-to-end temporal attention network (i.e., TAnet) was introduced in this work. TAnet employs a multi-head attention (MHA) mechanism, which can more effectively capture the interactions among time steps in collected EEG signals and efficiently assign corresponding weights to those EEG time steps. Experiments demonstrated that, compared with the CNN-based method and recent ASAD methods, TAnet provided improved decoding performance in the KUL dataset, with decoding accuracies of 92.4% (decision window 0.1 s), 94.9% (0.25 s), 95.1% (0.3 s), 95.4% (0.4 s), and 95.5% (0.5 s) with short decision windows (i.e., <1 s). As a new ASAD model with a short decision window, TAnet can potentially facilitate the design of EEG-controlled intelligent hearing aids and sound recognition systems.

*Keywords—auditory spatial attention detection, multi-head attention, electroencephalography*

## I. Introduction

The ability to focus on a specific speaker or sound source in a noisy environment has been visualized in the "cocktail party effect" [1]. Studies have shown that auditory attention is regulated by neural activity in the brain [e.g., 2-4]. Thanks to advanced brain imaging techniques, many studies are now able to decode the target of auditory attention in complex scenes involving many speakers.

Studies have shown that the brain encodes auditory spatial attention lateralized when the target and competing auditory streams are spatially separated [e.g., 5-7]. Researchers have proposed a series of auditory spatial attention detection (ASAD) models based on this finding. Current ASAD methods can be categorized into two main groups: rule-based methods, which usually consist of conventional decoders including feature extraction and pattern classification, and deep neural network-based methods. Most rule-based ASAD methods rely heavily on second-order statistics [e.g., 8-10], and the inherent nonlinear relationship between the listener's auditory attention and EEG paterns limits their detection accuracy. The performance of these methods degrades significantly under shorter decision windows. Since ASAD is a nonlinear classification problem, deep learning has been recognized as promising to provide better solutions [e.g., 11-17].

Vandecappelle et al. [11] proposed a convolutional neural network (CNN)-based ASAD model with a decoding accuracy of about 82% for a 1-second decision window in the KUL dataset [12]. Recent studies introduced neural networks based on attention mechanisms [e.g., 13-16]. STAnet [13] dynamically assigned discretized weights to EEG channels through spatial attention mechanisms. By temporally setting discretized weights to EEG signals on the decision window, STAnet achieved a decoding accuracy of 90.1% for a 1-second decision window in the KUL dataset [12]. Zhang et al. [14] recently proposed a learnable spatial mapping method incorporating a spatial attention mechanism to improve the effective utilization of the spatial distribution of EEG electrodes. XAnet [15] divided the EEG-based neural activities into left and right channel groups, and integrated them into cross-attention, yielding a 90.6% decoding accuracy for a 1-second decision window. EEG-Graph Net [16] achieved a 96.1% decoding accuracy for a 1-second decision window by modeling the topology of the human brain through the spatial patterns of the EEG signals as a graph.

Even on time scales as short as a few seconds, existing models can accurately decode the sound streams listeners attend to from their EEG data [e.g., 11-17]. However, most existing studies reported ASAD performance with decision windows ranging from 1 to 5 seconds, whereas this work attempts to further improve the decoding performance with a short decision window (i.e., in the 0.1 to 0.5 seconds range). This is because a short ASAD decision window is more helpful in capturing the dynamic changes in application scenarios that require fast and accurate recognition of sound streams. Previous studies suggested that brain activity is a temporal process and that human response to speech stimuli is dynamic and temporal [e.g., 18-19]. This means that considering the temporal dimension is critical in understanding the complex relationship between EEG signals and speech stimuli. Therefore, this study proposes a temporal attention mechanism that precisely captures the temporal features among EEG signals. This mechanism is similar to how the human brain selectively focuses on the incoming audio stimuli and dynamically assigns cognitive weights to the different EEG time steps.

This work was supported by the National Key Research and Development Program of China (Grant No. 2023YFF1203502), Guangdong Basic and Applied Basic Research Foundation (Grant No. 2022B1515120056), and the Basic Research Foundation of Shenzhen (Grant No. JCYJ20220818101217037).

This work aims to improve the ASAD accuracy under conditions with a short decision window (i.e., <1 s). An end-to-end temporal attention neural network (i.e., TAnet) is introduced, which utilizes a multi-head attention (MHA) [20] mechanism to capture critical features in the collected EEG signals, thereby further improving the decoding performance of ASAD within a short time window, especially from 0.1 to 0.5 seconds. The effectiveness of the temporal attention mechanism with a short decision window is verified through experiments in the KUL dataset [12] and performance comparison with several recent ASAD methods.

## II. METHODS

TAnet is an end-to-end architecture, as shown in Fig. 1, that combines a multi-head temporal attention module (MHTAM) and pooling and fully connected (FC) layers to automatically learn and recognize critical features directly from the raw EEG data, avoiding the need for traditional manual feature engineering.

### A. Multi-Head Temporal Attention Module

The core of TAnet is its MHTAM, as shown in Fig. 1(a), which can accurately capture and enhance important features characterizing auditory attention. The module receives raw multichannel EEG data $E \in \mathbb{R}^{T \times C}$, where $T$ represents the time samples, and $C$ represents the number of channels. MHTAM processes the input data $E$ through two independent attention heads, and each head searches for attention in a different representation subspace of the EEG data. The output of MHA [20] is then added to the original input data $E$ to form a residual connection, which helps avoid the problem of gradient extinction in deep model training. This is immediately followed by a layer normalization to normalize the results after the addition operation to ensure the stability of the training process. After that, the processed signal passes through two dense layers, and the first is followed by a Relu activation function that increases the nonlinearity, allowing the model to capture more complex features. After one more layer of normalization operation, the final output is denoted as $E'$ in Fig. 1(a).

The introduction of the MHTAM in TAnet is based on the complexity and multidimensionality of EEG data. This mechanism enhances the capability of feature representation as it allows the model to process and attend to multiple aspects of the input data simultaneously, thus capturing the features of the EEG signal more comprehensively. In addition, the multi-attention mechanism is also critical for understanding the complex interactions among different time steps in the EEG signal, allowing TAnet to analyze the brain's activity patterns more accurately. It also allows the model to learn and adapt dynamically by emphasizing the more critical or informative parts of the input data through known weights. This means that TAnet can flexibly adjust its focus to the most essential features in real-time in the changing EEG signal. Finally, MHA [20] integrates the output from each head to form a comprehensive feature representation. This representation contains information about each time step and integrates the interrelationships among time steps, providing a comprehensive multidimensional feature view for subsequent analysis and decision-making.

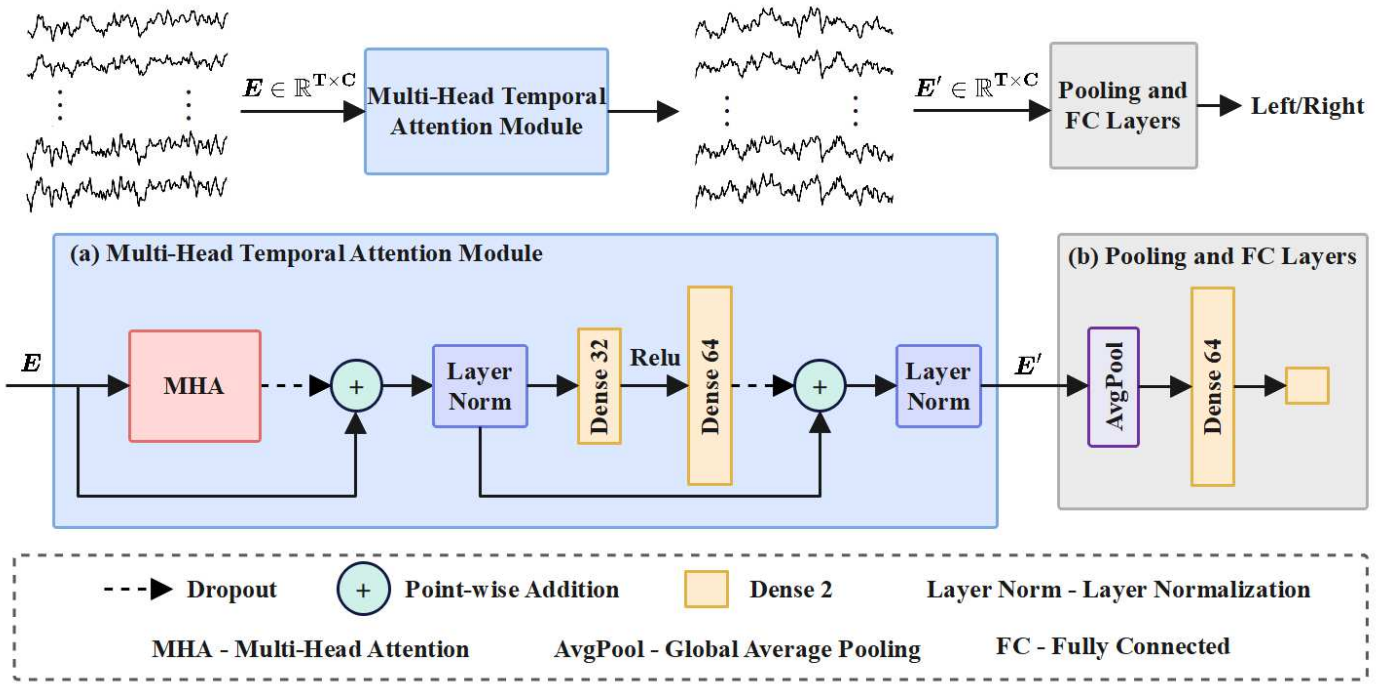

Fig. 1. A schematic diagram of the proposed temporal attention network, TAnet. TAnet mainly consists of (a) a multi-head temporal attention module and (b) a classification module. The network is trained on EEG data and detects auditory spatial attention through binary decision-making.

## B. Pooling and FC Layers

As shown in Fig. 1(b), TAnet employs a global average pooling operation applied to $E'$ after MHTAM. It reduces the dimensionality of the data while retaining the most critical information. This step is crucial to reduce computational complexity and prevent overfitting. After pooling, the data passes through a series of FC layers. These layers map the high-dimensional feature space to a lower-dimensional space where auditory spatial attention can be decoded more efficiently. The FC layers also introduce additional nonlinearities that improve the network's ability to model the complex relationships within the data.

## III. DATA AND EXPERIMENTS

TAnet is applied to auditory attention detection using a well-established dataset, i.e., the KUL dataset [12]. The results are compared with the baseline CNN method [11] and other best-performing approaches [e.g., 13, 16].

## A. Dataset

The KUL dataset, recorded by KULeuven [12], includes EEG data from 16 normal-hearing participants. Participants were instructed to focus on a designated speaker while ignoring simultaneous speech from another speaker. The speech stimuli consisted of four Dutch stories narrated by three male Flemish speakers. All stimuli underwent standardization to ensure consistent root mean squared (RMS) intensities, maintaining a uniform loudness perception. The stimuli were delivered in two ways: either dichotically, with one speaker assigned to each ear, or they underwent head-related transfer function (HRTF) filtering, replicating speech originating from 90 degrees to the listener's left and 90 degrees to the listener's right. Briefly, the subjects in the KUL dataset received different auditory stimuli in the left and right ears, respectively, rather than the mixed stimuli in [21]. EEG data were recorded from 64 channels using a BioSemi ActiveTwo device at a sampling rate of 8,192 Hz within an electromagnetically shielded and soundproof room. The KUL dataset, in total, comprises 72 minutes of EEG data per subject from 20 trials, accumulating 19.2 hours of EEG data across 16 subjects for experimental use. Additional details about the KUL dataset can be found in [e.g., 11-12].

## B. Data Preprocessing

For the KUL dataset, the EEG data from each channel were first re-referenced to the average response of that electrode. Since the analyzed EEG signals were collected at different sampling frequencies, they were all band-pass filtered through an FIR band-pass filter between 1 and 50 Hz and down-sampled to a sampling rate of 128 Hz. Next, the EEG data channels were z-score normalized to ensure that the mean of each trial was 0 and the variance was 1. Since the proposed TAnet is a data-driven solution expected to function end-to-end, the data processing process does not involve artifact removal operations. The simplified end-to-end process dramatically facilitates the realization of real-time application scenarios, such as neurally oriented hearing aids. In terms of decision window selection, five lengths of decision window, i.e., 0.1, 0.25, 0.3, 0.4, and 0.5 seconds, were analyzed to comprehensively validate the performance of TAnet with much shorter decision windows. Finally, because of the small sample size of EEG data for each subject and the heterogeneity of the data, sliding windows were applied to the EEG data to increase the diversity of the data and the generalization ability of the model. We focused on avoiding data leakage when generating the training and test sets. Specifically, we judged each training window and test window based on their left and right boundaries. We deleted the training window if there was a partial overlap to ensure no data duplication between the training and test sets.

## C. Experimental Setup

Five-fold cross-validation (CV) was utilized to evaluate the performance of TAnet and CNN baseline [11] across each decision window. The neural network was implemented using the TensorFlow framework, with the adaptive moment estimation (Adam) optimizer utilized to minimize the cross-entropy loss function. The learning rate was set to $10^{-3}$, the number of epochs was 300, and the batch size was 32. The patience parameter was set to 20 for the early stopping mechanism, ensuring a balanced approach to model training and validation.

In the KUL dataset [12], ASAD is a binary categorization task where decoding accuracy is the most direct and important performance metric. In order to simplify the evaluation process and to focus on the main goal of ASAD, using only accuracy as a metric makes the results easier to understand and interpret.

## IV. RESULTS AND DISCUSSION

In the auditory spatial attention decoding experiments performed in the KUL dataset [12], as Table I shows, TAnet performs excellently at all decision windows. Specifically, with a decision window of 0.1 s, TAnet achieves an impressive decoding accuracy of 92.4%. The decoding accuracy continues to rise, reaching 94.9%, 95.1%, 95.4%, and 95.5% for decision windows of 0.25 s, 0.3 s, 0.4 s, and 0.5 s, respectively.

In comparison, the accuracy of STAnet [13] with the 0.1 s decision window is only 80.8%. Although its accuracy increased to 87.2% with the 0.5 s decision window, it remained 8.3% lower than that of TAnet in this work. EEG-Graph Net [16] showed an accuracy of 88.7% with the 0.1 s decision window and reached 94.2% with the 0.5 s decision window, representing 3.7% and 1.3% lower accuracies than TAnet, respectively. EEG-Graph Net [16] is the current model with the highest decoding accuracy with short decision windows, but TAnet still has a significant performance improvement at 0.1 s decision window.

It is worth noting that, despite employing the same data processing procedure as TAnet, the CNN [11] baseline model falls short in decoding accuracy compared to TAnet across all decision windows. The CNN decoding accuracies at 0.1 s and 0.5 s decision windows were 78.9% and 89.0%, respectively. Figure 2 shows each participant's TAnet and CNN decoding accuracies at 0.1 s and 0.5 s decision windows. It further highlights the superiority of the TAnet-based ASAD performance with a short decision window.

TABLE I.  AUDITORY ATTENTION DETECTION ACCURACY (%) COMPARISON OF DIFFERENT MODELS IN THE KUL DATASET FOR DECISION WINDOWS OF DIFFERENT LENGTHS.

| Model | Length of decision window (second) | | | | |
|---|---|---|---|---|---|
| | 0.1 | 0.25 | 0.3 | 0.4 | 0.5 |
| STAnet [13] | 80.8 | - | - | - | 87.2 |
| EEG-Graph Net [16] | 88.7 | - | - | - | 94.2 |
| CNN [11] | 78.9 | 80.8 | 86.7 | 88.1 | 89.0 |
| **TAnet (This work)** | **92.4** | **94.9** | **95.1** | **95.4** | **95.5** |

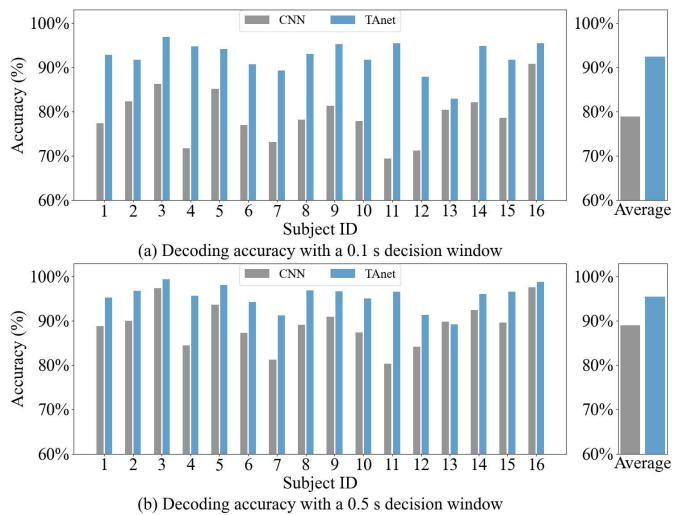

Fig. 2. ASAD accuracy of CNN [11] and TAnet. Panels (a) and (b) show the results for each subject in the KUL dataset using decision windows of 0.1 s and 0.5 s, respectively. The average accuracy for all subjects under two decision windows is shown on the right.

## V. CONCLUSION

This work proposed a new temporal attention network (i.e., TAnet) for EEG-based auditory spatial attention decoding. TAnet performs well in ASAD, especially with short decision windows. Its multi-head attention mechanism effectively captures the temporal features and dynamic interactions among EEG signals, making TAnet to stand out under tasks requiring a short decision window. The significant decoding accuracy improvement achieved in decision windows of 0.1 s to 0.5 s highlights the potential impact of TAnet in real-time applications such as intelligent hearing aids and fast sound recognition systems.